\documentclass[useAMS,usenatbib]{mn2e}
\usepackage{graphicx,lscape,subfigure}
\usepackage{latexsym}
\usepackage{epsfig}
\usepackage{Times}

\begin{document}
 
\title[Dark Bursts population in a complete sample of bright LGRBs]{The Dark Bursts population in a complete sample of bright {\it Swift} Long Gamma-Ray Bursts}


\author[A. Melandri et al.]{A. Melandri$^{1}$\thanks{E-mail: andrea.melandri@brera.inaf.it}, B. Sbarufatti$^{1}$, P. D'Avanzo$^{1}$, R. Salvaterra$^{2}$, S. Campana$^{1}$, S. Covino$^{1}$,  \newauthor S. D. Vergani$^{1,3}$, L. Nava$^{4}$, G. Ghisellini$^{1}$, G. Ghirlanda$^{1}$, D. Fugazza$^{1}$, V. Mangano$^{5}$, \newauthor M. Capalbi$^{6}$ \& G. Tagliaferri$^{1}$\\
$^{1}$INAF - Osservatorio Astronomico di Brera, via Bianchi 46, I-23807 Merate (LC), Italy\\
$^{2}$INAF - IASF Milano, via E. Bassini 15, I-20133 Milano, Italy\\
$^{3}$GEPI - Observatoire de Paris Meudon, 5 Place Jules Jannsen, F-92195 Meudon, France\\
$^{4}$SISSA, via Bonomea 265, I-34136 Trieste, Italy\\
$^{5}$INAF - IASF Palermo, via Ugo La Malfa 153, I-90146 Palermo, Italy\\
$^{6}$ASI Science Data Center, via G. Galilei, I-00044 Frascati (RM), Italy}

\label{firstpage} 

\maketitle

\begin{abstract} 

\noindent We study the properties of the population of optically dark events present in a carefully selected complete sample of bright {\it Swift} long gamma-ray bursts. The high level of completeness in redshift of our sample (52 objects out of 58) allow us to establish the existence of a genuine dark population and we are able to estimate the maximum fraction of dark burst events ($\sim 30 \%$) expected for the whole class of long gamma-ray burst. The redshift distribution of this population of dark bursts is similar to the one of the whole sample. Interestingly, the rest-frame X-ray luminosity (and the de-absorbed X-ray flux) of the sub-class of dark bursts is slightly higher than the average luminosity of the non-dark events. At the same time the prompt properties do not differ and the optical flux of dark events is at the lower tail of the optical flux distribution, corrected for Galactic absorption. All these properties suggest that dark bursts events generate in much denser environments with respect to normal bright events. We can therefore exclude the high-{\it z} and the low-density scenarios and conclude that the major cause of the origin of optically dark events is the dust extinction.

\end{abstract} 

 
\section{Introduction} 

Gamma Ray Bursts (GRBs) are brief and intense flashes of high energy gamma rays, originating at cosmological distances and often associated with radiation emitted at longer wavelengths for longer periods, identified as the afterglow. The afterglow is almost always detected in the X-ray band (for $\sim 95 \%$ of the events detected by {\it Swift}-XRT; Evans et al. 2009) while the optical afterglow is not. The GRBs that have no optical afterglow or a very low optical-to-X-flux ratio are classified as ``dark burst". Jakobsson et al. (2004) proposed that a GRB~should be classified as ``dark" if the slope of the spectral energy distribution between the optical and the X-ray band ($\beta_{\rm OX}$) is $< 0.5$. This working definition is a direct implication of the simplest fireball model. In fact, the spectral index $\beta$ ($F_{\nu} \propto \nu^{-\beta})$ is related to the power-law index of the electron energy distribution ($p$) and the location of the cooling frequency ($\nu_{\rm c}$), independently of the nature of circumburst environment (homogeneous or wind-like) and the collimation of the outflow (see Sari et al. 1998). The average value of $\beta_{\rm OX}$ is then expected to be between 0.5 (if $p$=2 and  $\nu_{\rm c}$ lies above the X-ray band) and 1.25 (if $p$=2.5 and  $\nu_{\rm c}$ lies below the optical frequency). Therefore any optically sub-luminous burst should be located below a constant line $\beta_{\rm OX} = 0.5$ in an optical vs. X-ray flux ($f_{\rm O}-f_{\rm X}$) diagram, providing that the fluxes are estimated at a common time, chosen to be $t_{\rm obs} = 11$~hr post-burst (Jakobsson et al. 2004). A slightly more elaborated method was presented by Rol et al. (2005) that compared the optical and X-rays fluxes at a given time extrapolating the latter to the optical band using not only the spectral index but also the temporal power-law index, in the context of the standard fireball model. Both Jakobsson et al. (2004) and Rol et al. (2005) found similar results ($\sim 10-20\%$ of possible dark bursts) on samples of pre-{\it Swift} GRBs and these methods are still used as immediate diagnostic tools to discriminate between optically bright and dark bursts.

Recently, van der Horst et al. (2009) proposed a new method for the optical classification of dark GRBs. Their method is less affected by assumptions about the emitting region with respect to the one done in Jakobsson et al. (2004). They improved the previous method by defining the region of optically (sub-luminous) dark bursts in the $\beta_{\rm OX}-\beta_{\rm X}$ plane: they are located below the dividing line of $\beta_{\rm OX} = \beta_{\rm X}-0.5$. In the same diagram, optically bright bursts are placed above the line $\beta_{\rm OX} = \beta_{\rm X}$, while all the GRBs that are still consistent with the fireball model will lie in the region defined by the relations $\beta_{\rm OX} = \beta_{\rm X}$ and $\beta_{\rm OX} = \beta_{\rm X}-0.5$. Optically bright events, for which the optical luminosity is too high if compared to the X-ray luminosity, are pretty rare, while there is a sizable fraction of events for which the X-ray emission seems to be in excess with respect to the observed optical one.

Based on these definitions, studies on GRBs samples in the {\it Swift}-era showed that, despite an advancement on the GRB~detection quality both in the prompt response and position accuracy, the fraction of genuinely dark GRBs remains significant. Melandri et al. (2008), Cenko et al. (2009), Zheng et al. (2009), Gehrels et al. (2008), Fynbo et al. (2009) and Greiner et al. (2011) found a fraction of dark GRB~in their samples of about 50\%, 50\%, 20\%, 20\%, 30\% and 40\% respectively. A higher fraction of dark GRBs is found in samples based on observations done by single ground based telescopes (i.e. the 2-m Liverpool and Faulkes Telescopes, Melandri et al. 2008; the 60-inch Palomar telescope, Cenko et al. 2009; the 2.2-m GROND telescope, Greiner et al. 2011) while a smaller fraction is detected if the whole {\it Swift} sample is considered (Gehrels et al. 2008 for GRBs up to end of 2007). This difference is certainly related to the different properties of the GRBs samples considered. In any case the population of dark bursts seems to be $\geq 20\%$ of the entire GRB~class.

In the era of rapid follow-ups the darkness of these events could not be ascribed to lack of sensitivity, late observational times or rapid temporal decays (Roming et al. 2006, Melandri et al. 2008). Different scenarios have been proposed to explain it: 
\begin{itemize}
\item {\it low-density scenario}: if the relativistic ejecta decelerate in a uniform low-density medium then the optical afterglow can be intrinsically faint with respect to the X-ray emission;
\item {\it dust scenario}: if dark bursts are exploding in galaxies with local thick and dusty (i.e., high metallicity) environments (with possibly some intervening systems along the line of sight) their optical afterglows could be suppressed by extinction, without thereby affecting their higher energy radiation. The extinction law characterising the bursts environment might be similar to the one observed in the local Universe or biased toward large dust grains;
\item {\it high redshift scenario}: if the burst is occurring at very large distances, its visible light could be completely extinguished as a result of the absorption of the Ly-$\alpha$ forest and Ly-$\alpha$ dropout redshifted into the optical bands. 
\end{itemize}
The latter explanation seems to be responsible for the non-detection only for a small fraction of the population of ``dark burst" (Greiner et al. 2011), while a combination of the first two effects seems to be a more realistic scenario (Perley et al. 2009).

In this paper, we will investigate the properties of the population of ``dark burst" present in a complete sub-sample of {\it Swift} long GRBs, with a high percentage of redshift determination (Salvaterra et al. 2011, Nava et al. 2011). The use of a complete sub-sample of GRBs allowed us to draw more firm conclusions about the properties of this class of events with no bias in the selection criteria. The text is organised as follows: in Section 2 we will describe the general properties of the dark bursts that belong to our selected sample. We then discuss the results on the dark bursts population, their redshift distribution and their luminosity in Section 3 and finally we draw our conclusions in Section 4. Throughout the paper we assume a standard cosmology with with $H_{0} = 70$~km~s$^{-1}$~Mpc$^{-1}$, $\Omega_{m} = 0.3$ and $\Omega_{\Lambda} = 0.7$.

\section{Sample selection}

The 58 GRBs in our sample have been selected to be relatively bright in the 15-150 keV {\it Swift}-BAT band, i.e. with the 1-s peak photon flux $P \ge $~2.6 ph s$^{-1}$ cm$^{-2}$, and have favourable conditions for ground-based multi-wavelength follow-up observations\footnote{ In particular we required that: i) the burst has been well localised by {\it Swift}-XRT and its coordinates quickly distributed; ii) the Galactic extinction in the burst direction is low, A$_{\rm V} < 0.5$; iii) the GRB declination is $-70^{\circ} < \delta < 70^{\circ}$; iv) the Sun-to-field distance is $\theta_{Sun} > 55^{\circ}$ and v) no nearby bright star is present.} (Salvaterra et al. 2011, Nava et al. 2011). This corresponds to an instrument that is $\sim$6 times less sensitive than {\it Swift}. Therefore, whatever GRB would have exploded in the sky with a flux equal or brighter than this limit, BAT would have detected for sure (if it was in its FOV). With this limit, no GRB would have been missed, in the meantime this value gives us also a reasonable number of GRB to perform statistical studies. Therefore our sample is complete with respect to this flux limit, it is of course biased toward the bright GRBs, but it is complete. Moreover, it turned out that $\sim 90\%$ of these bursts have also a redshift determination ($\sim 95\%$ have a constrained redshift).

From the observed light curve of each burst in our sample (as reported in the Burst Analyser of Evans et al. 2010) we estimated the fluxes at t$_{\rm obs}$=11 hr in the X-rays ($f_{\rm X}$ at 3 keV\footnote{We take into account all the effects due to X-ray absorption, even if at this energy they are negligible, and so $f_{\rm X}$ is the de-absorbed flux.}). In the optical band we used all the available public data to build the optical light curve, corrected for the Galactic absorption, and measure the optical flux ($f_{\rm O}$ in the $R$ filter) at t=t$_{\rm obs}$. We then used those two fluxes to calculate the values of the spectral index $\beta_{\rm OX}$ and we collected from the Spectrum Repository the values of the X-ray spectral index ($\beta_{\rm X}$) from the late time spectrum fit (Evans et al. 2009). All values are reported in Table \ref{tabDB}. For the majority of the bursts in our sample we were able to estimate the fluxes with good accuracy: in only one case (GRB~070328) it was not possible to estimate the optical flux due to the lack of optical observations. Instead, for the cases that had not enough detections to sample their decay, the value at 11 hr was estimated by interpolations and extrapolations of their observed light curve. We report below the assumptions that we made for these cases:
\\
- GRB~060814, GRB~061222A, GRB~070306, GRB~070521, GRB~080613B, GRB~090201, GRB~100621A: for these bursts we considered, as a conservative upper limit for $f_{\rm O}$, the closest (and deepest) upper limit (or detection) in the optical band to t=t$_{\rm obs}$. For these GRBs the individual times at which $f_{\rm O}$ was estimated were 0.97, 0.54, 33.7 (host detection), 0.61, 10.7, 7.5 and 6.0 hr, respectively. This allow us to put a safe upper limit on $\beta_{\rm OX}$; \\
- GRB~060306, GRB~061021:  for these bursts there were only few optical detections or upper limits so we needed to extrapolate $f_{\rm O}$ from the closest observation to t=t$_{\rm obs}$, assuming $\alpha_{\rm R}$ = 1.0; \\
- GRB~080603B: the last X-ray observation was acquired $\sim$ 3.3 hr after the burst. For this event we extrapolated $f_{\rm X}$ assuming the observed decay slope $\alpha_{\rm X} \sim$ 1.8;  \\
- GRB~060904A: for this event there is a gap in the XRT data that does not allow us to constrain the value of $\beta_{\rm OX}$. Using late time data acquired with the Photon Counting mode (PC, mean photon time arrival $\sim 58.5$~ks) we obtain $\beta_{\rm X}^{PC}$=$0.28^{+0.40}_{-0.47}$ while using the data acquired with the Windowed Timing mode (WT, mean photon time arrival $\sim 0.35$~ks) the value is $\beta_{\rm X}^{WT} = 1.07 \pm 0.05$. Moreover, there are no secure optical detections. Therefore, we could only put a conservative upper limit on the value of $f_{\rm O}$ for this event. Due to this uncertainty we excluded this event from our analysis;\\ 
- GRB~080602: only an optical upper limit at early time ($\sim 3.4$ hr post burst, Malesani et al. 2008) is available for this event. Also the XRT observations stop after $\sim$0.4 hr with a not well defined decay slope. This prevented us from estimating $f_{\rm O}$,  $f_{\rm X}$ and thus $\beta_{\rm OX}$, with good accuracy. We excluded also this event from our analysis;\\ 
- GRB~081221: as a conservative upper limit for the optical flux at t=t$_{\rm obs}$, we considered the only optical detection at $\sim 9$ hr, having in mind that this value could be contaminated by the host galaxy (Alfonso et al. 2008); \\
- GRB~090709A: the optical decay of this event is not well defined since the afterglow has been detected in the $R$ filter for only three epochs (Guidorzi et al. 2009, Cenko et al. 2010). We extrapolated the optical flux using the observed value of the optical decay ($\alpha_{\rm R} \sim$ 0.3) and we decided to consider this flux as conservative upper limit of $f_{\rm O}$;\\
- GRB~100615A: very conservatively we assumed the observed upper limit at $\sim 0.3$~hr (Nicuesa et al. 2010) as the upper limit for the optical flux. Even assuming the optical flux at such an early time, the $\beta_{\rm OX}$ remains pretty low ($\sim$ 0.06). In fact, the nature of this burst has already been analysed and discussed in detail by D'Elia et al. (2011), showing how this event is indeed a very dark burst.\\ 

After this analysis we ended up with a total of 55 GRBs (49 with secure redshift) for which it was possible to estimate the value (or an upper limit) of $\beta_{\rm OX}$ at the observed time t$_{\rm obs}$=11 hr post-burst.

\begin{figure}
   \centering
   \includegraphics[height=8.5cm,width=8.5cm]{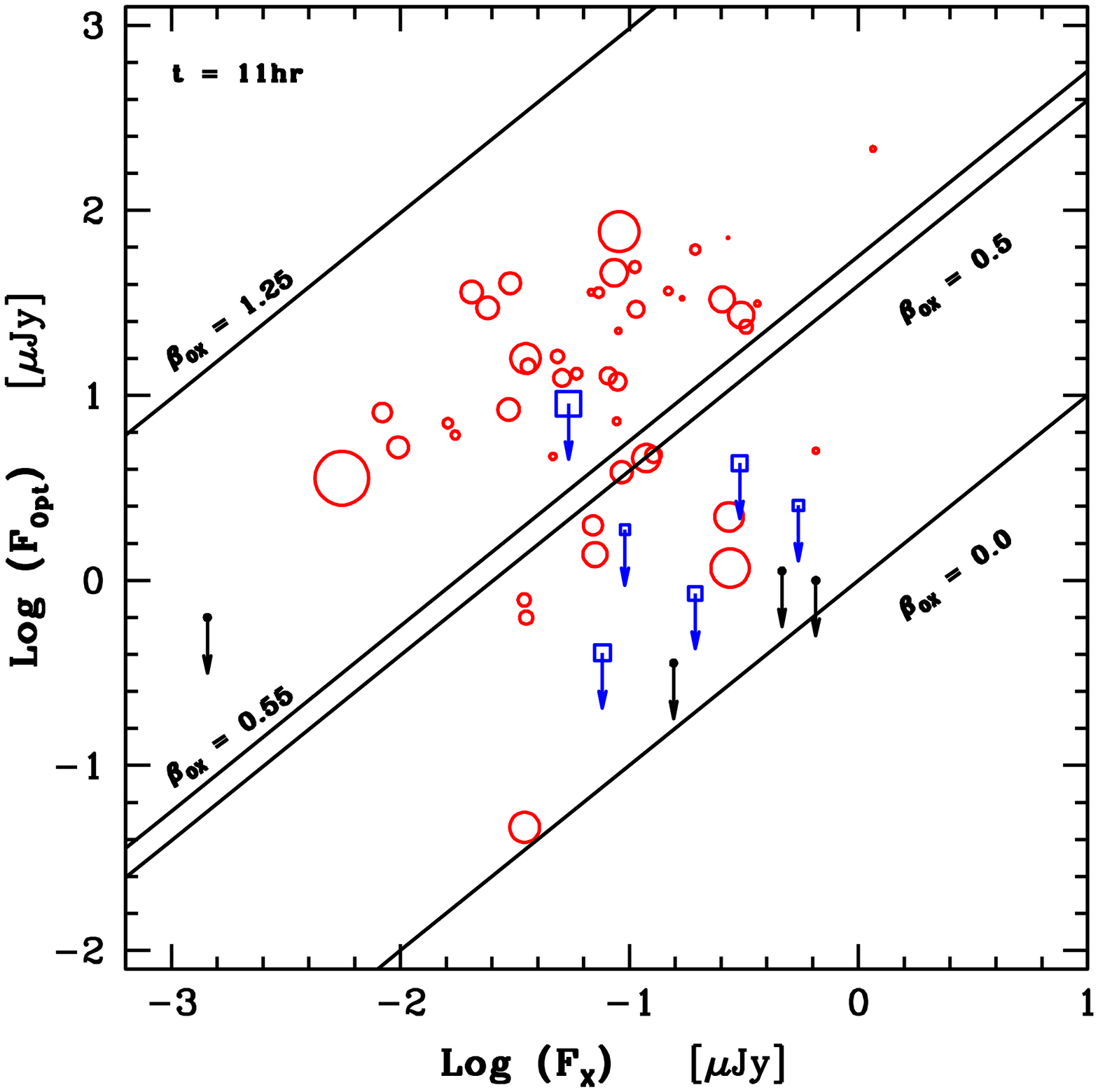} 
   \includegraphics[height=8.5cm,width=8.5cm]{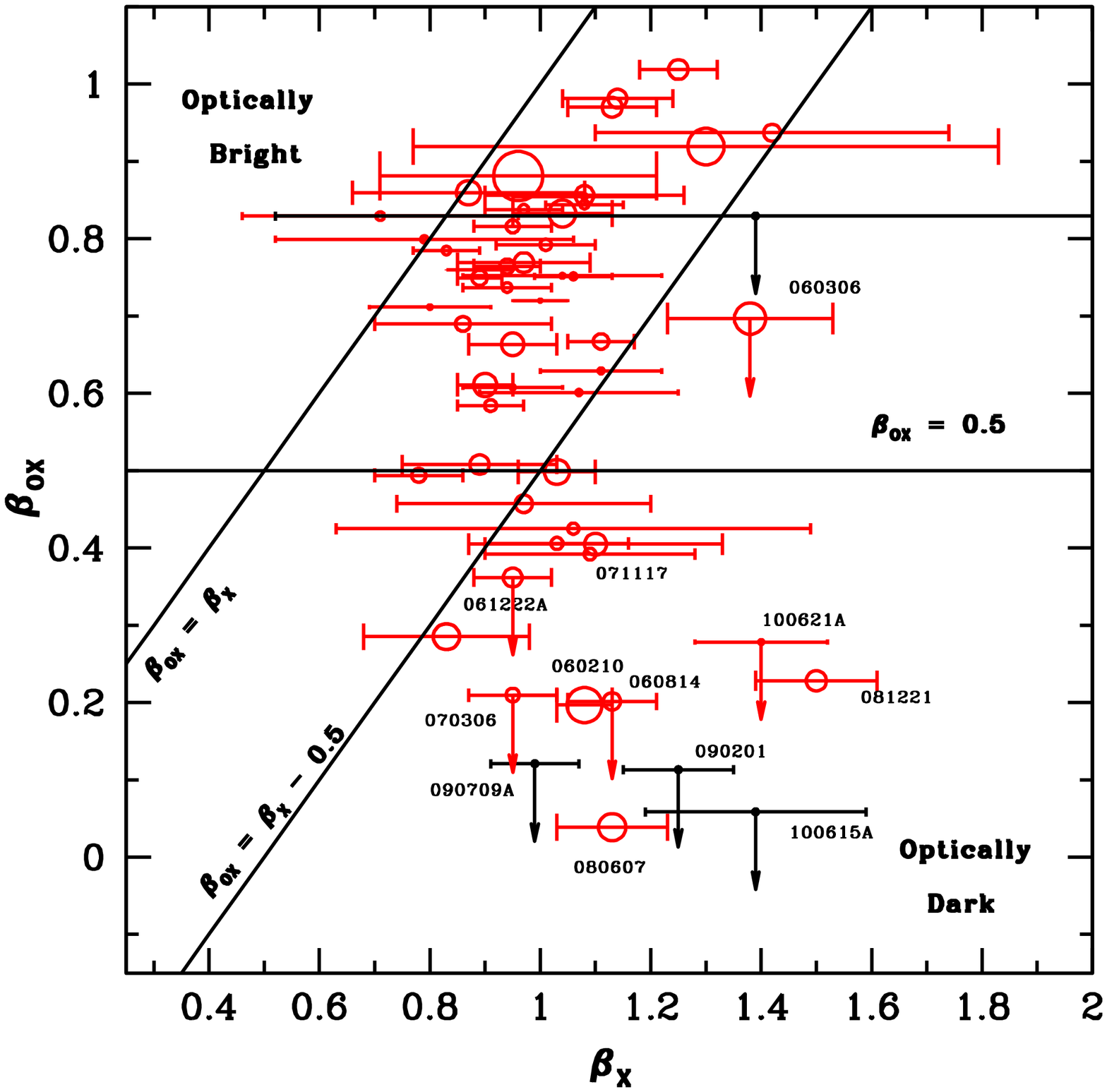} 
   \caption{{\it Top panel}: dark bursts distribution in our sample according to the definitions of Jakobsson et al. (2004). Open red circles and blue squares are GRBs with secure redshift, while filled black circles are the GRBs with no redshift measurement. {\it Bottom panel}: dark bursts distribution according to the definitions of van der Horst et al. (2009). The dimension of the symbol for both plots is a direct visual of the value of the redshift of the GRB, the larger the symbol the bigger the associated redshift.}
      \label{fig1}
\end{figure}

\section{Results and Discussion} 

\subsection{Dark Bursts Population}

For all the GRBs in our sample we estimated the optical flux ($f_{\rm O}$) and the X-rays flux ($f_{\rm X}$) in the observed frame at a common time t$_{\rm obs} = 11$~hr. Then we calculated the values of $\beta_{\rm OX}$ and we took the estimates of $\beta_{\rm X}$ from the {\it Swift} burst Spectrum Repository (late time PC-mode data, Evans et al. 2009). With these data we reproduced the dark bursts distribution of our sample according to the definition and diagram of Jakobsson et al. (2004) and van der Horst et al. (2009). The results are shown in Fig. \ref{fig1}. 

Following the practical definition of Jakobsson et al. (2004) we find a total of 18 GRBs lying below the $\beta_{\rm OX} = 0.5$ line: 10 of them are optically detected (9 with secure redshift: GRB~050401, GRB~060210, GRB~071117, GRB~080319C, GRB~080607, GRB~081222, GRB~090102, GRB~090812 and GRB~090926B; 1 with no redshift: GRB~090709A), while for the remaining 8 only upper limits in the optical bands are available (5 with redshift: GRB~060814, GRB~061222A, GRB~070306, GRB~070521 and GRB~100621A; 3 with no redshift: GRB~081221, GRB~090201 and GRB~100615A). In the $\beta_{\rm OX}-\beta_{\rm X}$ plane defined by van der Horst et al. (2009) only 11 out of these 18 events above (GRB~060210, GRB~060814, GRB~061222A, GRB~070306, GRB~071117, GRB~080607, GRB~081221, GRB~090201, GRB~090709A, GRB~100615A and GRB~100621A) still fall into the region for the optically dark events\footnote{One further event (GRB~060306) falls into the $\beta_{\rm OX}<\beta_{\rm X}-0.5$ region. However, there are large uncertainties in the extrapolation at t$_{\rm obs}$ of the optical flux and we decided not to include this event in the list of secure dark bursts. We note that there also some indication from $K$-band observations (Lamb et al. 2006) in favour of the dark nature of this object, therefore we include this event when we estimate the maximum fraction of dark bursts for the van der Horst et al. (2009) definition.}. The remaining 7 (GRB~050401, GRB~080319C, GRB~081222, GRB~090102, GRB~090812, GRB~090926B and GRB~050721) still have $\beta_{\rm OX}$ consistent with $\beta_{\rm X}-0.5$.

\setcounter{table}{1}
\begin{table} 
\begin{center}
 \caption{Fraction of dark bursts in our sample according to the definitions of Jakobsson et al. (2004) and van der Horst et al. (2009). The third column represents the strongest upper limit of the fraction of dark bursts when considering also the three excluded GRBs as possible dark bursts.}  \label{tabstat} \small
 \begin{tabular}{@{}ccc}
\hline 
\hline
Definition & Dark Bursts & Max Dark Bursts \\
 & \% & \% \\
\hline
$\beta_{\rm OX} < 0.5$ & 32.7 & $<36.2$  \\
$\beta_{\rm OX} < \beta_{\rm X} - 0.5$ & 20.0 & $<25.9$\\
\hline
\hline
 \end{tabular}
\end{center}
\end{table}

The fraction of dark bursts in our sample, including both detections and optical upper limits, is $\sim 32.7\%$ (18 out of 55 events) according to the Jakobsson et al. (2004) definition and $\sim 20.0\%$ (11/55) with respect to the van der Horst et al. (2009) diagram. These results are similar to previous studies of dark bursts in the {\it Swift}-era (Gehrels et al. 2008; Fynbo et al. 2009) confirming the existence of a genuine dark bursts population. Finally, if we include in our analysis also the three GRBs for which we do not have an accurate estimate of $\beta_{\rm OX}$ (events in italic font in Table \ref{tabDB}) we obtain a strong upper limit for the fraction of the population of dark bursts of $\sim$ 36\% in the case of Jakobsson et al. (2004) definition; the upper limit is $\sim$ 26\% in the case of the van der Horst et al. (2009) definition, for which we consider also GRB~060306 as a possible dark (Table \ref{tabstat}). 

We note that from our analysis GRB~050401 is not classified as dark in the van der Horst et al. (2009) diagram, while these authors classified it as dark. This difference is due to the fact that van der Horst et al. (2009) used the quick available X-ray spectral index $\beta_{\rm X}$, while we decided to use the more accurate value from the late time spectrum. This choice should be more accurate as the average values of $\beta_{\rm X}$ is usually estimated around the chosen time for the measure of the optical and X-ray fluxes.

\subsection{Darkness evolution}

The historical choice to extrapolate the optical and X-ray fluxes to the common time t$_{\rm obs} = 11$~hr has been motivated by the need of measuring only the radiation arising from the afterglow component, ensuring the cessation of the possible central engine activity and the end of the plateau phase. However, the nature of a dark burst can be further investigated by looking at the evolution of its darkness. In principle, early and/or late time central engine activity can mask the real forward shock X-ray emission, adding an additional component that might be not so relevant at later times. Therefore the total flux in that band at early time would be higher than the expected emission from the X-ray forward shock alone. The increase of the X-ray emission with respect to the optical one, for example during the so-called plateau phase, would change the value of $\beta_{\rm OX}$ and therefore the estimate of the darkness for some events (previously noted also in Melandri et al. 2008). Flares are seen only for few events in our sample and they are not responsible for the darkness of the events in our sample. We investigated the darkness evolution of the GRBs in our sample by estimating, when possible, the spectral index $\beta_{\rm OX}^{\prime}$ at an earlier time t$^{\prime}_{\rm obs} = 600$~s after the burst event. 

\begin{figure}
   \centering
   \includegraphics[height=8.5cm,width=8.5cm]{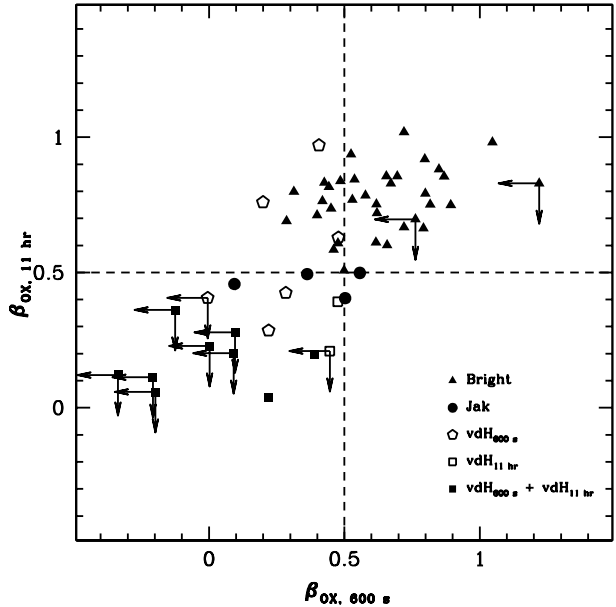} 
   \caption{Darkness evolution from t$^{\prime}_{\rm obs} = 600$~s to t$_{\rm obs} = 11$~hr of the events in our sample. See legend and main text for details about different symbols; ``Jak" and ``vdH" refer to Jakobsson et al. (2004) and van der Horst et al. (2009) dark burst definition, respectively.}
      \label{figbeta}
\end{figure}

We then reclassified the dark bursts according to the definition of van der Horst et al. (2009) and compared the values of $\beta_{\rm OX}$ at early and late times. Results are shown in Fig. \ref{figbeta}: this diagram is divided in four regions differently populated. The upper left quadrant is occupied by bright events (filled triangles) and by those events that are classified dark at early time but they are not dark at t$_{\rm obs} = 11$~hr (open symbols): typical example of this class is GRB~050416A, classified as dark at t=600~s, still dark at t=1000~s (Cenko et al. 2009, Perley et al. 2009) but no longer dark at late time. These are the events for which the central engine is probably still active at early time. In the upper right quadrant are located the GRBs that are always bright. The bottom right quadrant is the region that would be populated by GRBs that are not dark at early time and that evolve to become dark at late time: this is an implausible case and this region is indeed not populated. This region would be populated by bursts having an additional X-ray component at late time. Finally, in the bottom left quadrant we find those events that are classified as dark according to van der Horst et al. (2009) at any given time (filled squares), the events that are compatible with the criterium of van der Horst et al. (2009) at t=t$_{\rm obs}$ or at t=t$^{\prime}_{\rm obs}$ (open symbols), and also the remaining dark bursts according to the Jakobsson et al. (2004) definition only (filled circles). 

We note that, according to the definition of Jakobsson et al. (2004), the fraction of dark bursts at early time ($\sim 50\%$) is much higher than the fraction at late time ($\sim 33\%$, Table \ref{tabstat}), while with the criterium of van der Horst et al. (2009) these fractions are similar, being $\sim 27\%$ and $\sim 20\%$ at early and late time, respectively. This is clearly visible in Fig. \ref{figbeta} where some bright events populate the upper left quadrant. Such a high percentage of dark bursts was previously reported in works based on sample of GRBs observed with ground based facilities (Melandri et al. 2008, Cenko et al. 2009). In general we can say that bona fide dark events are the ones that are found to be dark both at early and late time ($\geq 14\%$, filled squares in Fig. \ref{figbeta}). This fraction increases to a maximum of $\sim 25-35\%$ when considering only late time optical and X-ray emission.

\subsection{Dark Bursts redshift distribution}

\begin{figure}
   \centering
   \includegraphics[height=8.5cm,width=8.5cm]{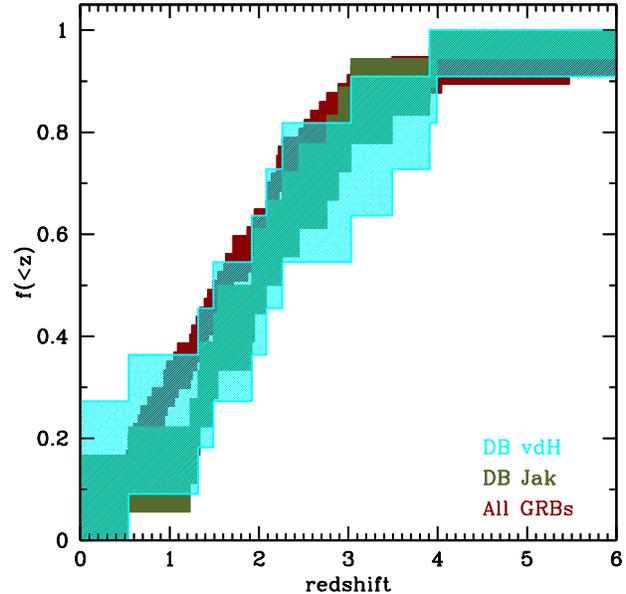} 
   \caption{Cumulative redshift distribution of dark bursts (cyan and dark olive according to van der Horst et al. (2009) and Jakobsson et al. (2004) definition, respectively) compared with our whole sample (dark red).}
      \label{figz}
\end{figure}

Thanks to the high completeness in redshift of our sample we built the cumulative redshift distribution for the sub-class of optically dark bursts (Fig. \ref{figz}). In this plot we show the distribution for our entire sample of GRBs together with the ones for dark bursts according to both definitions from literature (``DB Jak" for Jakobsson et al. 2004; ``DB vdH" for van der Horst et al. 2009). To quantify the existence of a separated populations of optically dark bursts, we applied the Kolmogorov-Smirnov statistic to our sub-sample of optically dark events. 

For both definitions we compared the distribution of dark bursts population to one of the whole sample, including ({\it z$_{\rm all}$}) and excluding ({\it z$_{\rm no-DB}$}) the dark bursts considered. Results are reported in Table \ref{KStDB}, where we quantify the maximum deviation between the cumulative redshift distributions (D) and the associated probability that two set of data are drawn from the same distribution (P). In order to say something conclusive about two populations being separated the value of P should be as lower as possible. Clearly, the sub-sample of dark bursts, independently by the definition, it is consistent with coming from the same population of the whole sample (Table 3). 

The range of redshift for the events in our sample belonging to the optically dark region spans from 0.54 up to 3.91. The contribution of high-{\it z} events in our sample for the dark bursts population can be estimated to be $\leq 3.6 \%$; in fact, only 2 events out of the 55 that we considered in our analysis do not have a redshift determination and can, in principle, be at very high-{\it z}. Theoretical models for GRBs redshift distribution predict $\leq 1$ dark event at a redshift $z > 6$ (Salvaterra et al. 2011), that is indeed what we observe in our complete sample of 58 GRBs (Fig. \ref{figz}). This allow us to ascertain that the darkness of the fraction of bursts in our sample that satisfy the dark bursts definition of van der Horst et al. (2009) is not due to the so-called high redshift scenario. 

\setcounter{table}{2}
\begin{table} 
\begin{center}
 \caption{Result of the Kolmogorov-Smirnov tests for our sample. DB(Jak) are the events classified as dark according to the definition of Jakobsson et al. (2004), while DB(vdH) are the dark events following van der Horst et al. (2009).}  \label{KStDB}
 \begin{tabular}{@{}ccc}
\hline 
Samples & D & P \\
\hline
\hline
$z_{\rm DB(Jak)}$ vs. $z_{\rm all}$ & 0.062 & 0.999 \\
$z_{\rm DB(vdH)}$ vs. $z_{\rm all}$ & 0.122 & 0.999 \\
$z_{\rm DB(Jak)}$ vs. $z_{\rm no-DB(Jak)}$ & 0.065 & 0.999 \\
$z_{\rm DB(vdH)}$ vs. $z_{\rm no-DB(vdH)}$ & 0.119 & 0.999 \\
$z_{\rm DB(Jak)}$ vs. $z_{\rm DB(vdH)}$ & 0.116 & 0.999 \\
\hline
Observed frame & & \\
\hline
f$_{\rm X, 11h, DB}$ vs f$_{\rm X, 11h, no-DB}$ & 0.517 & 0.010 \\
f$_{\rm O, 11h, DB}$ vs f$_{\rm O, 11h, no-DB}$ & 0.767 & 2.14 $\times 10^{-5}$ \\ 
\hline
Rest frame & & \\
\hline
L$_{\rm X, 11h, DB}$ vs L$_{\rm X, 11h, no-DB}$ & 0.461 & 0.076 \\
E$_{\rm iso, DB}$ vs E$_{\rm iso, no-DB}$ & 0.375 & 0.240 \\
E$_{\rm peak, DB}$ vs E$_{\rm peak, no-DB}$ & 0.225 & 0.840 \\
L$_{\rm iso, DB}$ vs L$_{\rm iso, no-DB}$ & 0.350 & 0.314 \\
\hline
\hline
 \end{tabular}
\end{center}
\end{table}

\begin{figure*}
   \centering
   \includegraphics[height=5.5cm,width=5.5cm]{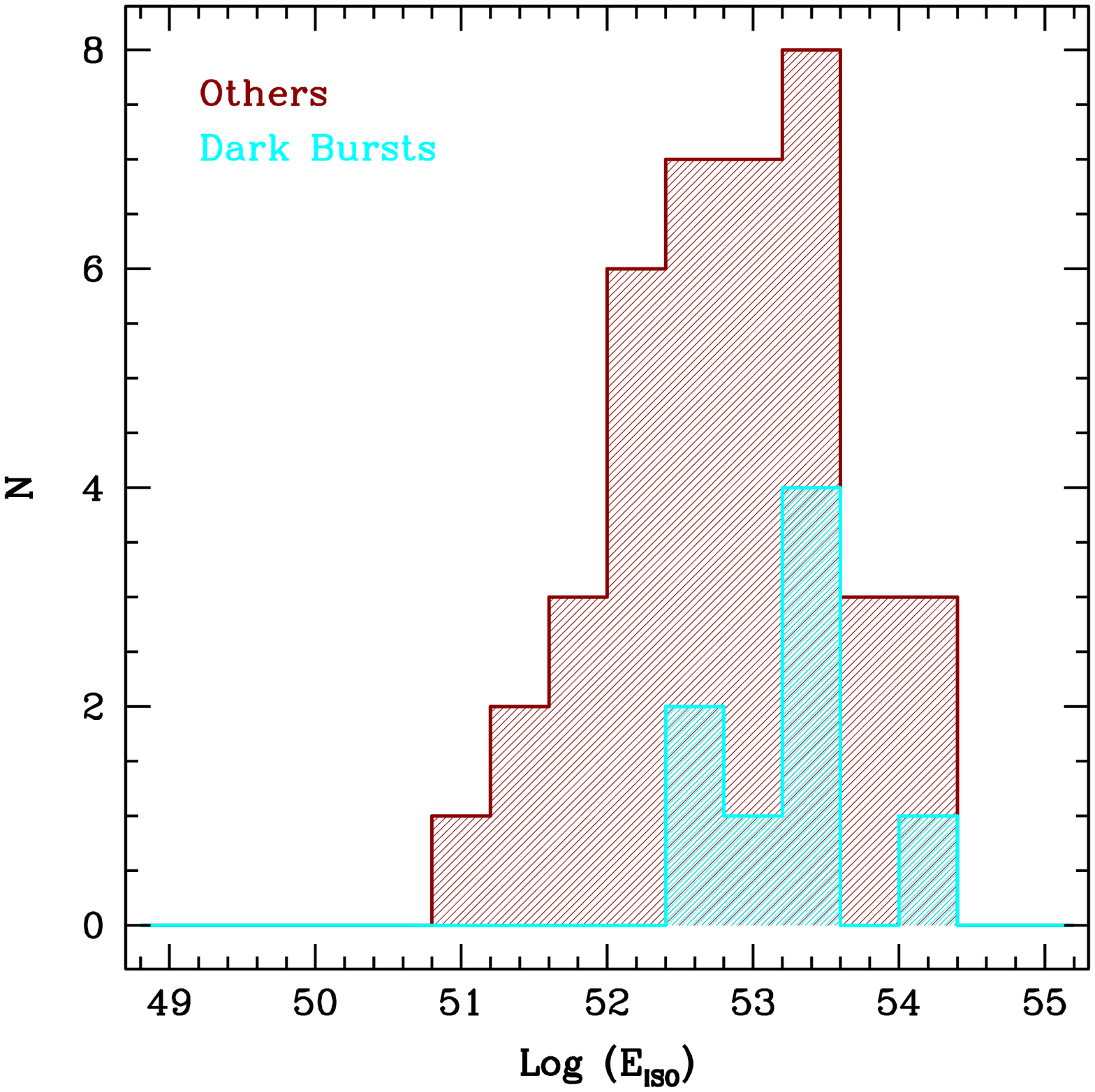} 
   \includegraphics[height=5.5cm,width=5.5cm]{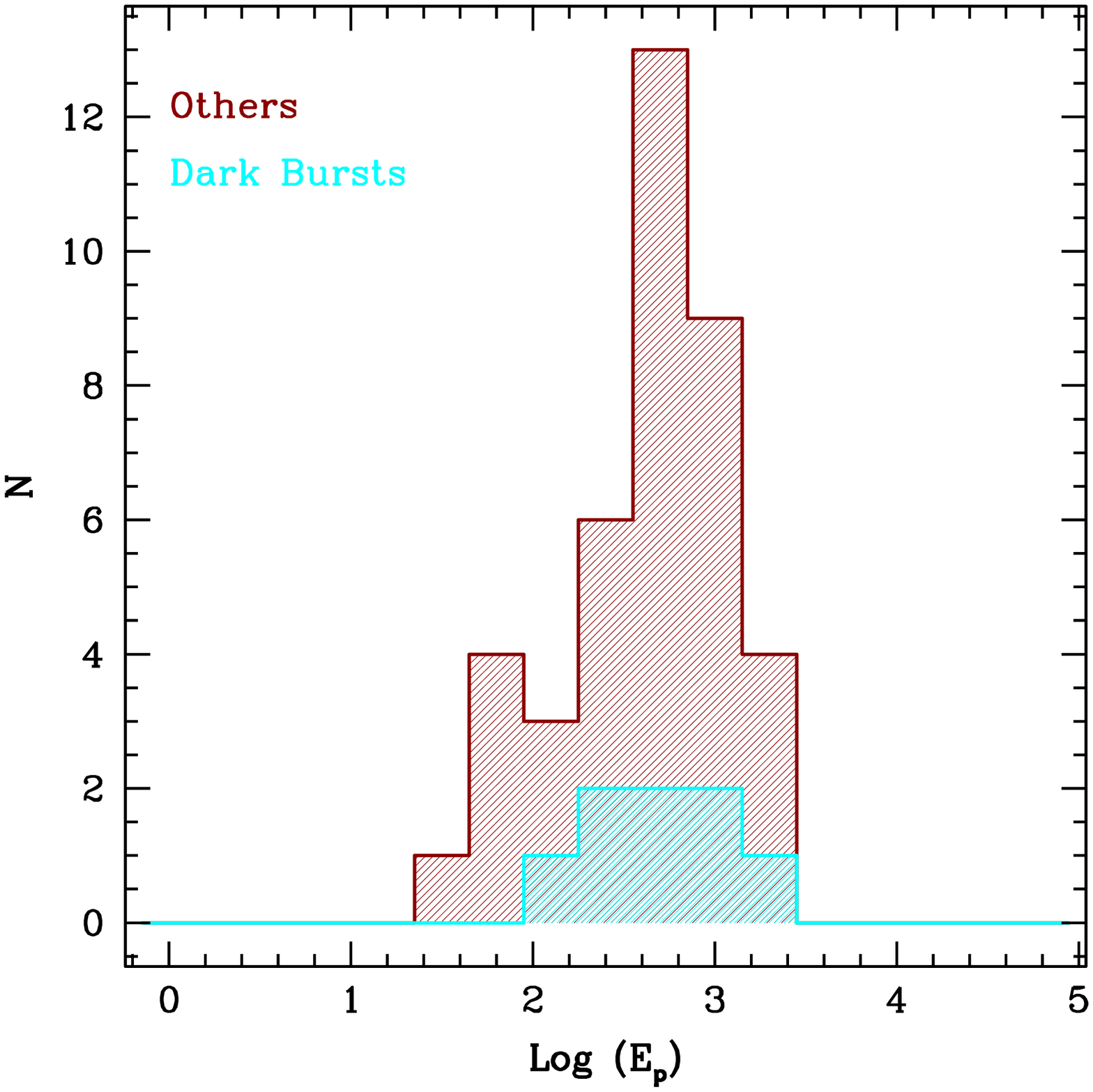} 
   \includegraphics[height=5.5cm,width=5.5cm]{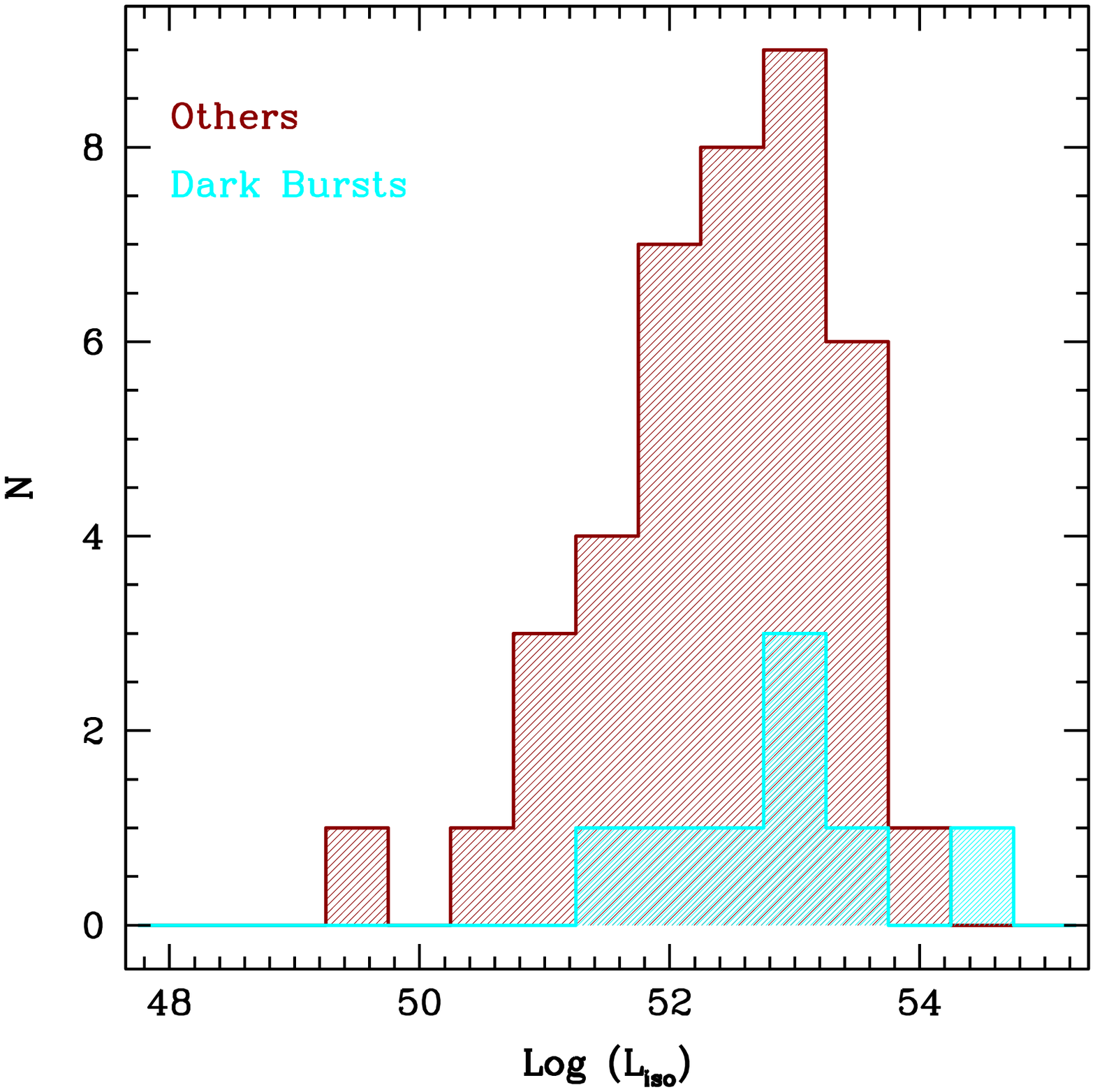} 
   \includegraphics[height=5.5cm,width=5.5cm]{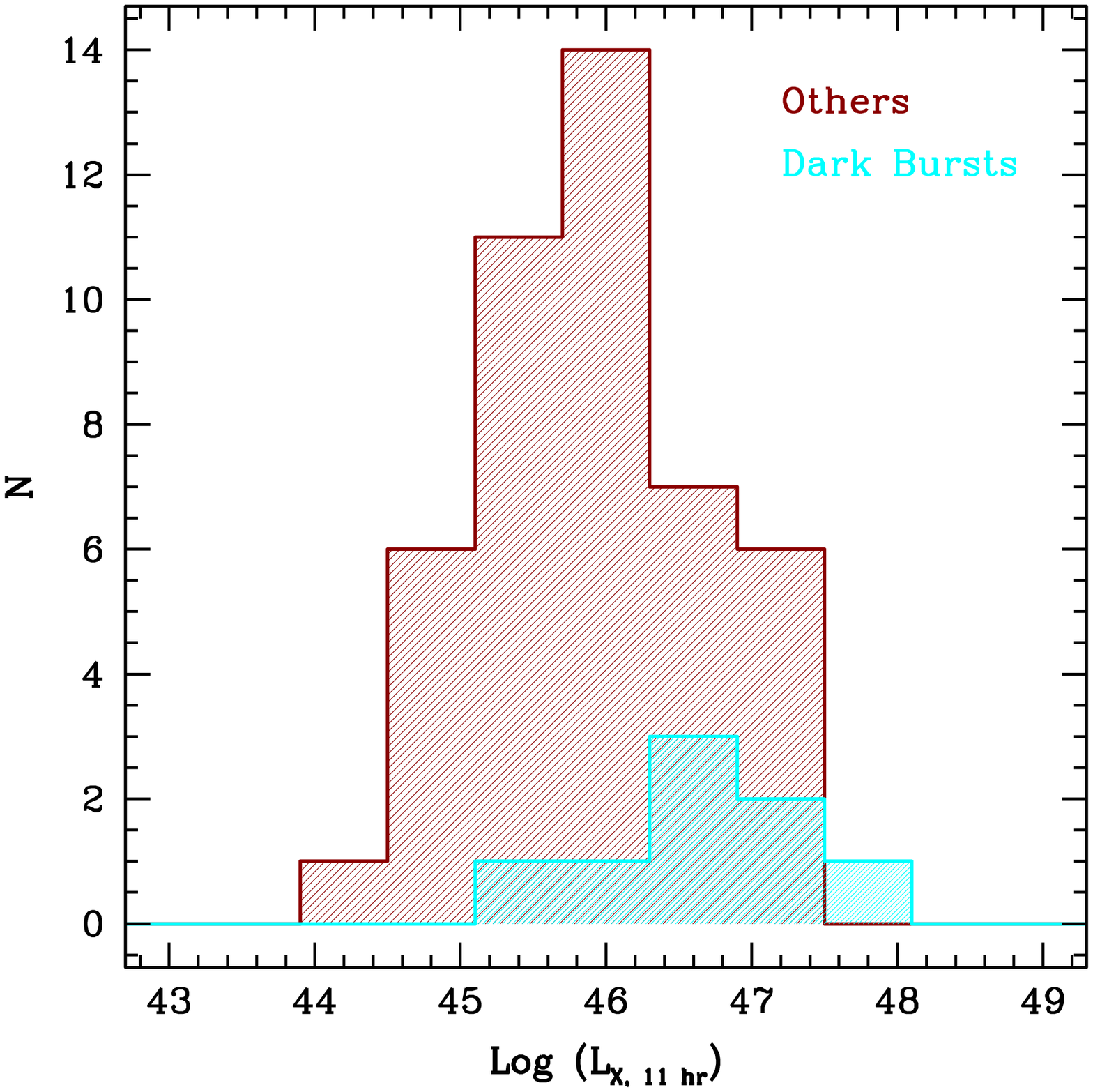} 
   \includegraphics[height=5.5cm,width=5.5cm]{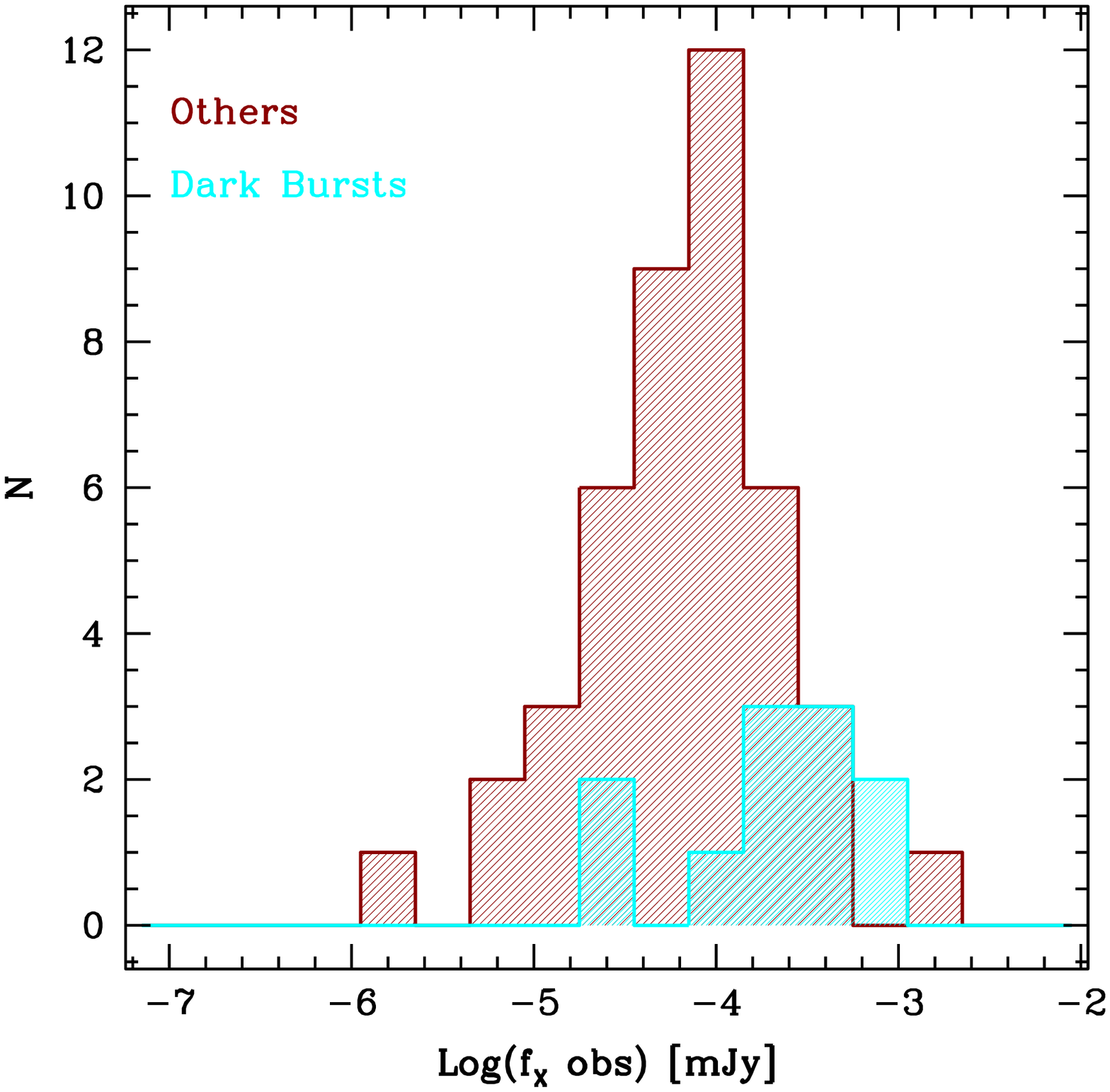} 
   \includegraphics[height=5.5cm,width=5.5cm]{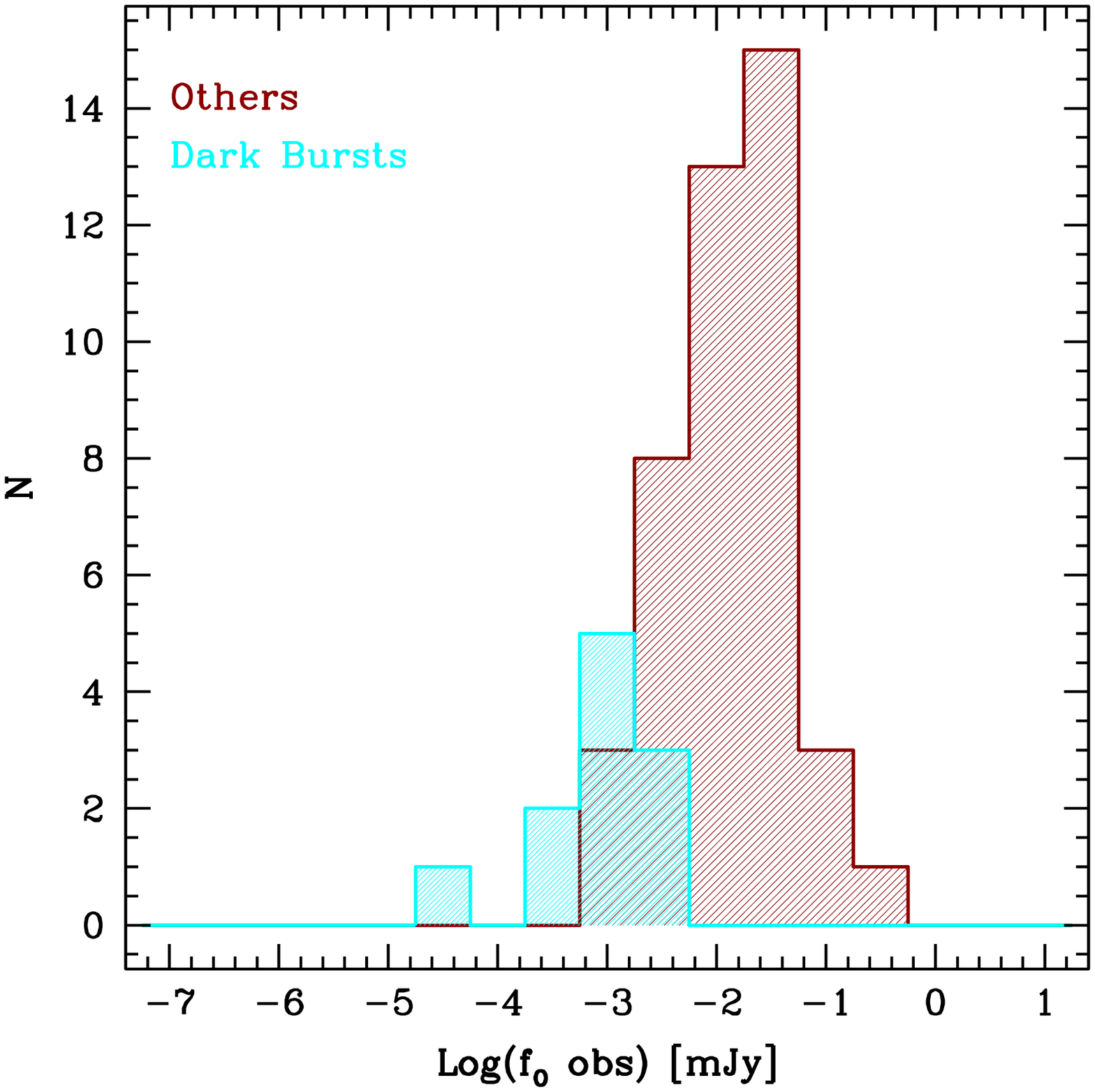} 
   \caption{Histograms of the properties of dark (cyan) and non-dark (dark red) events: dark bursts have similar prompt properties (here we show the estimated isotropic energy E$_{\rm iso}$, the peak energy E$_{\rm peak}$ and the isotropic luminosity L$_{\rm iso}$), higher X-ray luminosity (L$_{\rm X}$) and observed flux (f$_{\rm X}$), and lower observed optical flux (f$_{\rm O}$).}
      \label{figlumx}
\end{figure*}

\subsection{Dark Bursts X-ray Luminosity}

Using the observed X-ray fluxes we calculated the rest frame unabsorbed X-ray luminosity for each event with secure redshift in our sample at a common rest-frame time t = 11 hr. We then investigated the properties of the sub-class of dark bursts compared to the bright bursts.

Pre-{\it Swift} studies of the X-ray properties of optically bright and dark bursts shows that the latter seems to be on average 5 times fainter in the X-ray band than optically bright events (de Pasquale et al. 2003). In their sample, de Pasquale and collaborators had 31 GRBs, with 20 events belonging to the ``dark" sub-class and the remaining 11 bursts with an optically detected afterglow. However, their definition of ``dark" event was only based on the absence in the optical band of a detected afterglow, resulting in upper limits for the optical flux of these events. Using those upper limits to infer the upper limits on $\beta_{\rm OX}$ we find that only 3 events (of the 20 reported in their Table 1) would have been classified as dark bursts according to the definition of Jakobsson et al. (2004). Therefore their results are biased and not representative of the differences between the dark and bright populations.

Instead, in our complete sample we have an indication that the observed X-ray fluxes of the dark events are, on average, larger than the fluxes of bright events. This is still true, but slightly less significant (Table \ref{KStDB} for the results of the KS tests), when we consider the rest-frame X-ray luminosity. In Fig. \ref{figlumx} we show the histograms that summarise the properties of the dark population with respect to the standard bright bursts. The prompt properties, like E$_{\rm iso}$, E$_{\rm peak}$ and L$_{\rm iso}$, do not differ and are still consistent with a single population of events\footnote{We compared the prompt properties of dark and bright events in our sample using the values computed by Nava et al. (2011).}. The class of dark bursts do show a clear difference on the amount of X-ray absorbing column density, having systematically higher column densities with respect to bright events, as studied in detail by Campana et al. (2011). In the optical and X-ray bands the two classes seems to be well defined, although a clear dichotomy is not visible. The dark bursts are at the same time in the lower tail of the optical flux distribution and on the higher tail of the X-ray flux distribution. In other words they are consistently less luminous in optical and more luminous in X-rays. This is conserved and still visible in the high tail of the X-ray luminosity distribution (Fig. \ref{figlumx}).

The higher X-ray luminosity for some GRBs can be a direct consequence of the fact that the X-ray emission is the sum of emission coming from internal and external processes, long lived central engine or late prompt emission (i.e., Ghisellini et al. 2007). An additional emission will enhance the X-ray emission with respect to the one expected from the forward shock emission alone. In the case of the dark population in our sample, the contribution of one of these components might be more pronounced, resulting in a higher X-ray luminosity than the ``normal" events. However, this might be the explanation for those events that display an evolution of their darkness from early to late time, since this additional component might be more active at early times. As shown in Fig. \ref{figbeta} this is the case for few events that would be classified as dark at early time, that subsequently lost their darkness at later time (upper left quadrant in Fig. \ref{figbeta}). None of the GRBs classified as dark in our sample display this behaviour and therefore, even if they tend to lie in the high end of the X-ray flux and luminosity distribution, this cannot be the explanation of their darkness. 

A possible explanation of the slightly higher X-ray luminosity of the dark bursts could be found in the different local environment of these event with respect to normal bursts. Dark bursts are the events that display the higher X-ray column densities (Campana et al. 2011), indication of a metal-rich environment where the absorption is more efficient. In that case the $N_{\rm H}/A_{\it V}$ ratio might be significantly lower for these events, that for a fixed value of $N_{\rm H}$ translates into a higher value of $A_{\it V}$. Therefore the attenuation of the X-ray emission for dark bursts can be significantly lower with respect to the absorption in their optical band.

\section{Conclusions} 

We studied the properties of the sub-class of optically dark bursts detected in the complete sample of bright {\it Swift} long GRBs presented in Salvaterra et al. (2011) and Nava et al. (2011). From our analysis we find that this population has the same redshift distribution of the whole sample. Thanks to the high completeness in redshift of this sample we have been able to confirm the existence of a genuine fraction ($\sim 25-35 \%$) of optically dark events.  The majority of those do not show any darkness evolution, being optically dark from very early time. Those events cannot be explained in the context of to the high-{\it z} scenario and we confidently exclude that their darkness is due to the Ly-$\alpha$ absorption in the optical bands. 

The dark bursts do not have different prompt properties compared to the normal events (see Table \ref{KStDB} and Fig. \ref{figlumx}). However the former display lower optical flux and relatively higher X-ray flux with respect to the latter, as also noted by van der Horst et al. (2009). In particular they are always located in the high tail of the X-ray luminosity distribution, showing that they are, at the same time, not only fainter in the optical but also brighter in the X-ray. Because we are dealing with bright prompt events, it is unlikely that the darkness of the events in our sample could be ascribed only to their intrinsic faintness or to rapid temporal decay. These bursts are indeed faint events in the optical band but they also have a slight excess of emission in the X-ray band (although the excess is not statistically significant as the optical deficiency). The higher X-ray column densities observed for the dark bursts in our sample (Campana et al. 2011) clearly indicate that they formed in metal-rich environments where a fair amount of dust must be present. This disfavour the low-density scenario. 

Therefore the most plausible explanation is left to be found in the context of the dust scenario. Since we take into account the Galactic absorption, if this scenario is correct then the observed darkness is due to high local extinction in a dense environment or to absorption from intervening material. Unfortunately we do not have complete information regarding all the possible intervening systems that the light of these events may encounter along their lines of sight and probably this effect becomes relevant only for GRBs at high redshift (Campana et al. 2006, Campana et al. 2010). Therefore it can play a significant role only for a small fraction of dark bursts in our sample. The former possibility instead plays for sure an important role for a couple of events in our sample: GRB~060210 (A$_{\rm V} \geq 4$; Curran et al. 2007) and GRB~080607 (A$_{\rm V} \geq 3.5$; Perley et al. 2009).  

High values of X-ray column densities are hint of high local absorption; however, we do not know what happens to the dust environment in the vicinity of the GRB. Their prompt emission could, for example, alter the local dust composition, destroying small dust grains in favour of the bigger ones. This may change the extinction law that will become flatter, nearly constant and independent from the observed wavelength, in the UV-optical band, with respect to the one commonly observed in the local Universe. This effect, known as gray dust, even if it is not easy to recognise was successfully invoked to explain few GRB spectral energy distributions (Stratta et al. 2005). We do not have clear hints of this scenario in our sub-sample of dark bursts but we cannot exclude that it plays a significant role. However the investigation of the ``dust scenario" and the ``gray dust scenario" needs a more detailed analysis of the spectral energy distribution of each single event in our sample; this is beyond the aim of this work and will be exhaustively treated in a dedicated forthcoming work.

On the basis of our results we were able to understand more about the nature of dark bursts when compared to bright events:
\begin{itemize}
\item they have similar prompt properties;
\item they have a higher X-ray flux and X-ray luminosity and, at the same time, lower optical flux;
\item they are located in different (denser) environments;
\item they cannot be explained in the context of the high-{\it z} or low-density scenarios;
\item their darkness must be mainly related to circum-burst dust absorption. 
\end{itemize}

\section*{Acknowledgments} 

We thank the referee for providing constructive and well-directed comments. This work has been supported by ASI grant I/011/07/0. This work made use of data supplied by the UK {\it Swift} Science Data Centre at the University of Leicester.

\setcounter{table}{0}
\begin{landscape}
\begin{table} 
\begin{center}
 \caption{Properties of our samples: we reported the observed X-ray ($f_{\rm X}$) and optical ($f_{\rm O}$) fluxes at t =11 hr post burst together with the $\beta_{\rm X}$  from the {\it Swift} burst Spectrum Repository (Evans et al. 2009) and the estimated values of $\beta_{\rm OX}$ (at t$_{\rm obs} = 11$~hr) and $\beta_{\rm OX}^{\prime}$ (at t$^{\prime}_{\rm obs} = 600$~s.). Optical fluxes have been corrected for the Galactic absorption and obtained by: $^{a}$ interpolation of the light curve; $^{b}$ extrapolation of the light curve; $^{c}$ assuming as a conservative value the closest upper limit (the exact times are reported in Section 2). GRBs with $\beta_{\rm OX} < 0.5$ are in bold font, while in italic we show the three GRBs that we excluded from our analysis (see main text for details).}  \label{tabDB} \footnotesize
 \begin{tabular}{@{}ccccccc|ccccccc}
\hline
\hline 
GRB~& redshift & $f_{\rm X}$ & $f_{\rm O}$ & $\beta_{\rm X}$ & $\beta_{\rm OX}$ & $\beta_{\rm OX}^{\prime}$ & GRB~& redshift & $f_{\rm X}$ & $f_{\rm O}$ & $\beta_{\rm X}$ & $\beta_{\rm OX}$ & $\beta_{\rm OX}^{\prime}$ \\
  & & [$\mu$Jy] & [$\mu$Jy] & &  & & & & [$\mu$Jy] & [$\mu$Jy] & & &\\
\hline 
050318 & 1.44 & 0.036 $\pm$ 0.010 & 14.386 $\pm$ 1.633$^{b}$ & $0.95^{+0.07}_{-0.06}$ & 0.816 & 0.443 & {\it 080602} & $\sim$1.4 & | & $<$3.703$^{c}$ & $0.90^{+0.12}_{-0.13}$ & | & | \\                                                
{\bf 050401} & 2.90 & 0.272 $\pm$ 0.067 & 2.205 $\pm$ 0.210$^{a}$ & $0.83^{+0.15}_{-0.14}$ & 0.285 & 0.220 & 080603B & 2.69 & 0.085 $\pm$ 0.016 & 45.996 $\pm$ 6.373$^{a}$ & $0.87^{+0.26}_{-0.21}$ & 0.859 & 0.695 \\                
050416A & 0.65 & 0.046 $\pm$ 0.012 & 4.678 $\pm$ 0.296$^{a}$ & $1.11^{+0.11}_{-0.14}$ & 0.629 & 0.477 & 080605 & 1.64 & 0.080 $\pm$ 0.017 & 12.758 $\pm$ 0.470$^{a}$ & $0.86^{+0.11}_{-0.16}$ & 0.689 & 0.286 \\                 
050525A & 0.61 & 0.068 $\pm$ 0.014 & 36.076 $\pm$ 5.913$^{a}$ & $1.08^{+0.15}_{-0.13}$ & 0.854 & 0.868 & {\bf 080607} & 3.04 & 0.034 $\pm$ 0.008 & 0.046 $\pm$ 0.004$^{a}$ & $1.13^{+0.06}_{-0.11}$ & 0.038 & 0.217 \\            
050802 & 1.71 & 0.050 $\pm$ 0.014 & 12.417 $\pm$ 1.190$^{b}$ & $0.89^{+0.04}_{-0.07}$ & 0.749 & 0.893 & 080613B & | & 0.0014 $\pm$ 0.0007 & $<$0.631$^{c}$ & $1.39^{+1.28}_{-0.87}$ & $<$0.829 & $<$1.220 \\                         
050922C & 2.20 & 0.020 $\pm$ 0.005 & 36.194 $\pm$ 4.527$^{a}$ & $1.25^{+0.06}_{-0.07}$ & 1.019 & 0.720 & 080721 & 2.59 & 0.307 $\pm$ 0.064 & 27.216 $\pm$ 2.510$^{a}$ & $0.91^{+0.05}_{-0.05}$ & 0.611 & 0.616 \\                 
060206 & 4.05 & 0.090 $\pm$ 0.021 & 76.699 $\pm$ 3.943$^{a}$ & $1.30^{+0.57}_{-0.53}$ & 0.919 & 0.797 & 080804 & 2.20 & 0.029 $\pm$ 0.008 & 8.385 $\pm$ 0.309$^{a}$ & $0.97^{+0.12}_{-0.12}$ & 0.769 & 0.529 \\                  
{\bf 060210} & 3.91 & 0.275 $\pm$ 0.067 & 1.164 $\pm$ 0.291$^{a}$ & $1.08^{+0.05}_{-0.05}$ & 0.196 & 0.390 & 080916A & 0.69 & 0.088 $\pm$ 0.024$^{a}$ & 7.239 $\pm$ 0.200 & $1.07^{+0.13}_{-0.18}$ & 0.601 & 0.657\\                 
060306 & 3.50 & 0.054 $\pm$ 0.012 & $<$9.009$^{b}$ & $1.38^{+0.06}_{-0.15}$ & $<$0.696 & $<$0.762 & 081007 & 0.53 & 0.089 $\pm$ 0.023 & 22.315 $\pm$ 0.822$^{a}$ & $1.04^{+0.10}_{-0.18}$ & 0.752 & 0.618 \\                 
060614 & 0.13 & 0.269 $\pm$ 0.066 & 71.082 $\pm$  3.943$^{a}$ & $0.89^{+0.06}_{-0.04}$ & 0.759 & 0.199 & 081121 & 2.51 & 0.254 $\pm$ 0.050 & 33.012 $\pm$ 6.116$^{a}$ & $0.95^{+0.08}_{-0.08}$ & 0.663 & 0.792 \\                 
{\bf 060814} & 1.92 & 0.194 $\pm$ 0.050 & $<$0.851$^{c}$ & $1.13^{+0.07}_{-0.07}$ & $<$0.201 & $<$0.091 & 081203A & 2.10 & 0.030 $\pm$ 0.007 & 40.409 $\pm$ 6.507$^{a}$ & $1.14^{+0.09}_{-0.10}$ & 0.981 & 1.047\\                
{\it 060904A} & | & 0.079 $\pm$ 0.019 & $<$0.001$^{b}$ & $0.28^{+0.74}_{-0.47}$ & | & | & {\bf 081221} & 2.26 & 0.076 $\pm$ 0.021 & $<$0.406$^{c}$ & $1.50^{+0.12}_{-0.11}$ & $<$0.228 & $<$0.003 \\                 
060908 & 1.88 & 0.008 $\pm$ 0.002 & 8.079 $\pm$ 0.361$^{a}$ & $1.42^{+0.32}_{-0.36}$ & 0.937 & 0.524 & {\bf 081222} & 2.77 & 0.118 $\pm$ 0.031 & 4.586 $\pm$ 0.550$^{a}$ & $1.03^{+0.07}_{-0.06}$ & 0.498 & 0.557\\            
060912A & 0.94 & 0.016 $\pm$ 0.004 & 7.075 $\pm$ 2.786$^{a}$ & $0.71^{+0.19}_{-0.25}$ & 0.829 & 0.671 & {\bf 090102} & 1.55 & 0.127 $\pm$ 0.031 & 4.773 $\pm$ 0.484$^{a}$ & $0.78^{+0.06}_{-0.08}$ & 0.493 & 0.362\\            
060927 & 5.47 & 0.005 $\pm$ 0.002 & 3.568 $\pm$ 1.447$^{a}$ & $0.96^{+0.33}_{-0.25}$ & 0.881 & $0.642$ & {\bf 090201} & $<$4.0 & 0.155 $\pm$ 0.037 & $<$0.357$^{c}$ & $1.25^{+0.10}_{-0.10}$ & $<$0.113 & $<$-0.207 \\                
061007 & 1.26 & 0.048 $\pm$ 0.012 & 16.240 $\pm$ 0.228$^{a}$ & $1.01^{+0.09}_{-0.06}$ & 0.792 & 0.799 & 090424 & 0.54 & 0.361 $\pm$ 0.082 & 31.319 $\pm$ 2.309$^{a}$ & $0.95^{+0.10}_{-0.09}$ & 0.607 & 0.476 \\                 
061021 & 0.35 & 0.169 $\pm$ 0.047 & 33.468 $\pm$ 2.058$^{b}$ & $1.00^{+0.04}_{-0.05}$ & 0.720 & 0.620 & {\bf 090709A} & $<$3.5 & 0.463 $\pm$ 0.113 & $<$1.124$^{b}$ & $0.99^{+0.08}_{-0.08}$ & $<$0.120 & $<$-0.336\\               
061121 & 1.31 & 0.322 $\pm$ 0.079 & 23.474 $\pm$ 0.345$^{a}$ & $0.91^{+0.06}_{-0.06}$ & 0.584 & 0.460 & 090715B & 3.00 & 0.035 $\pm$ 0.010 & 15.853 $\pm$ 0.730$^{a}$ & $1.04^{+0.09}_{-0.09}$ & 0.832 & 0.425 \\                
{\bf 061222A} & 2.09 & 0.302 $\pm$ 0.082 & $<$4.303$^{c}$ & $0.95^{+0.07}_{-0.06}$ & $<$0.361 & $<$-0.125 & {\bf 090812} & 2.45 & 0.070 $\pm$ 0.018 & 1.381 $\pm$ 0.216$^{a}$ & $0.95^{+0.07}_{-0.06}$ & 0.404 & 0.503 \\            
{\bf 070306} & 1.50 & 0.545 $\pm$ 0.113 & $<$2.543$^{c}$ & $0.95^{+0.07}_{-0.06}$ & $<$0.209 & $<$0.446 & {\bf 090926B} & 1.24 & 0.035 $\pm$ 0.005 & 0.748 $\pm$ 0.117$^{b}$ & $0.95^{+0.07}_{-0.06}$ & 0.424 & 0.283\\            
{\it 070328} & $<$4.0 & 0.230 $\pm$ 0.058 & | & $0.95^{+0.08}_{-0.08}$ & | & | & 091018 & 0.97 & 0.073 $\pm$ 0.020 & 35.938 $\pm$ 6.657$^{a}$ & $1.10^{+0.18}_{-0.23}$ & 0.844 & 0.537\\                 
{\bf 070521} & 1.35 & 0.095 $\pm$ 0.020 & $<$1.878$^{c}$ & $1.03^{+0.15}_{-0.13}$ & $<$0.405 & $<$-0.005 & 091020 & 1.71 & 0.088 $\pm$ 0.024 & 11.866 $\pm$ 0.437$^{a}$ & $1.11^{+0.05}_{-0.06}$ & 0.667 & 0.720\\                 
071020 & 2.15 & 0.092 $\pm$ 0.031 & 3.843 $\pm$ 2.294$^{a}$ & $0.89^{+0.16}_{-0.14}$ & 0.508 & 0.499 & 091127 & 0.49 & 1.157 $\pm$ 0.284 & 214.595 $\pm$ 39.757$^{a}$ & $0.80^{+0.11}_{-0.11}$ & 0.711 & 0.399\\               
071112C & 0.82 & 0.017 $\pm$ 0.004 & 6.105 $\pm$ 0.561$^{a}$ & $0.79^{+0.21}_{-0.27}$ & 0.799 & 0.313 & 091208B & 1.06 & 0.058 $\pm$ 0.013 & 13.102 $\pm$ 3.665$^{a}$ & $0.94^{+0.13}_{-0.08}$ & 0.736 & 0.451\\                
{\bf 071117} & 1.33 & 0.035 $\pm$ 0.008 & 0.628 $\pm$ 0.080$^{a}$ & $1.09^{+0.13}_{-0.19}$ & 0.392 & 0.475 & {\bf 100615A} & | & 0.651 $\pm$ 0.156 & $<$1.000$^{c}$ & $1.39^{+0.20}_{-0.20}$ & $<$0.058 & $<$-0.198 \\                     
080319B & 0.94 & 0.195 $\pm$ 0.051 & 61.390 $\pm$ 4.772$^{a}$ & $0.82^{+0.06}_{-0.06}$ & 0.784 & 0.577 & {\bf 100621A} & 0.54 & 0.416 $\pm$ 0.098 & $<$5.021$^{c}$ & $1.40^{+0.13}_{-0.12}$ & $<$0.278 & $<$0.096 \\                 
{\bf 080319C} & 1.95 & 0.069 $\pm$ 0.016 & 1.984 $\pm$ 0.091$^{a}$ & $0.97^{+0.28}_{-0.23}$ & 0.457 & 0.094 & 100728B & 2.11 & 0.010 $\pm$ 0.002 & 5.242 $\pm$ 0.532$^{a}$ & $1.08^{+0.17}_{-0.18}$ & 0.856 & 0.655\\                 
080413B & 1.10 & 0.105 $\pm$ 0.028 & 49.461 $\pm$ 1.822$^{a}$ & $0.97^{+0.05}_{-0.07}$ & 0.838 & 0.485 & 110205A & 2.22 & 0.024 $\pm$ 0.005 & 29.684 $\pm$ 2.738$^{a}$ & $1.13^{+0.09}_{-0.09}$ & 0.970 & 0.406\\                
080430 & 0.77 & 0.148 $\pm$ 0.041 & 36.649 $\pm$ 2.567$^{a}$ & $1.06^{+0.06}_{-0.07}$ & 0.751 & 0.817 & 110503A & 1.613 & 0.106 $\pm$ 0.029 & 29.189 $\pm$ 2.422$^{a}$ & $0.95^{+0.04}_{-0.06}$ & 0.764 & 0.419 \\               
\hline
\hline
\hline
 \end{tabular}
\end{center}
\end{table}
\end{landscape}

\label{lastpage} 

\end{document}